\documentclass{elsart}
\usepackage{graphicx,amssymb,psfig}

\newcommand{\be}{\begin{equation}}
\newcommand{\ee}{\end{equation}}
\newcommand{\bea}{\begin{eqnarray}}
\newcommand{\eea}{\end{eqnarray}}

\begin{document}
\begin{frontmatter}

\title{Thermodynamical Observables in a Finite Temperature Window from the Monte Carlo Hamiltonian}

\author[Laval]{H. Kr\"{o}ger\thanksref{cor}},
\thanks[cor]{Talk given by H. Kr\"{o}ger}
\ead{hkroger@phy.ulaval.ca}
\author[Zhongshan]{X.Q. Luo},
\ead{stslxq@zsu.edu.cn}
\author[Dalhousie]{K.J.M. Moriarty}
\ead{moriarty@cs.dal.ca}
 
\address[Laval]{D\'{e}partement de Physique, Universit\'{e} Laval, 
Qu\'{e}bec, Qu\'{e}bec G1K 7P4, Canada. } 

\address[Zhongshan]{Department of Physics, Zhongshan University, 
Guangzhou 510275, China. } 
 
\address[Dalhousie]{Department of Mathematics, Statistics and 
Computer Science, Dalhousie University, Halifax, Nova Scotia B3H 3J5, 
Canada }

\begin{abstract}
The Monte Carlo (MC) Hamiltonian is a new stochastic method to solve many-body problems. The MC Hamiltonian represents an effective Hamiltonian in a finite energy window. We construct it from the classical action via Monte Carlo with importance sampling. The MC Hamiltonian yields the energy spectrum and 
corresponding wave functions in a low energy window. This allows to compute thermodynamical observables in a low temperature window.  
We show the working of the MC Hamiltonian by an example from lattice field theory (Klein-Gordon model).
\end{abstract}
\begin{keyword}
Many-body systems, thermodynamical observables, Monte Carlo methods
\end{keyword}
\end{frontmatter}

\section{Introduction}
\label{sec:Introduction}
Many interesting phenomena in physics occur in many-body systems.
The theoretical difficulty lies in solving models of many-body systems.
A brilliant idea is the renormalisation group approach 
\`a la Kadanoff and Wilson \cite{Kadanoff}. It suggests to compute 
critical phenomena by constructing an effective (renormalized) Hamiltonian, by thinning out degrees of freedom in an appropriate manner.
Here we suggest to adapt this idea for the purpose to construct 
an effective Hamiltonian in a low-energy window.
The basic idea to thin out degrees of freedom is taken over from 
lattice field theory, where infinite-dimensional integrals 
(path integrals) are successfully simulated by a small 
number of representative configurations.
In analogy to the representative configurations of path integrals,
we construct a representative basis of Hilbert states 
via a Monte Carlo (MC) algorithm with importance sampling.  
Furthermore we use the MC algorithm to compute matrix 
elements of the transition amplitude.

In recent years, Monte Carlo methods have been widely used to solve problems in quantum physics. 
For example, with quantum Monte Carlo there has been improvement in nuclear shell model calculations \cite{Otsuka:01}. 
A proposal to solve the sign problem in Monte Carlo Greens function method, 
useful for spin models has been made by Sorello \cite{Sorello:98}.
Lee et al. \cite{Lee:01} have suggested a method to diagonalize Hamiltonians, 
via a random search of basis vectors having a large overlap with low-energy eigenstates. 
In contrast to that, in this work we construct matrix elements of the transition amplitude and hence the Hamiltonian via the path integral starting from the action.

\section{Monte Carlo Hamiltonian}
\label{sec:MCHamiltonian}
Let us discuss the construction of the MC Hamiltonian introduced in Ref.\cite{Jirari:99} in 
several steps. First, consider in 1-dim quantum mechanics the motion 
of a single particle of mass $m$ under the influence of a local potential 
$V(x)$. Its classical action is given by
\be
S = \int dt ~ \left[ \frac{m}{2} \dot{x}^{2} - V(x) \right] ~ .
\ee
Given the classical action, one can determine 
quantum mechanical (Q.M.) transition amplitudes.
Similar to the approach in Lagrangian lattice field theory, we use imaginary time in what follows. 
We consider the transition amplitude for some finite time $T$ ($t_{in}=0$, $t_{fi}=T$) and for all combinations of positions $x_{i}, x_{j}$.
Here $x_{i}$, $x_{j}$ run over a finite discrete set of points $\{x_{1},\dots,x_{N}\}$ located on the real axis. Suppose these points are equidistantly distributed (spacing $\Delta x$) over some interval $[-L,+L]$. 
The transition amplitudes are given by the (Euclidean) path integral,
\begin{equation}
\label{eq:TransAmpl}
M_{ij}(T)
= \int [dx] \exp[ - S_{E}[x]/\hbar ]\bigg |_{x_{j},0}^{x_{i},T} ,              
\end{equation}
where $S_{E}$ denotes the Euclidean action.
Suppose for the moment that those transition amplitudes were known. 
From that one can construct an approximate, i.e. effective Hamiltonian.
Note that the matrix $M(T)=[M_{ij}(T)]_{N \times N}$ is a real, positive and Hermitian matrix. It can be factorized into a unitary matrix $U$ 
and a real diagonal matrix $D(T)$, 
\be
\label{eq:Factor}
M(T)=U^{\dagger}~D(T)~U ~ .
\ee
The matrix $M(T)$ can be expressed in terms of the full Hamiltonian $H$
by
\be
M_{ij}(T) = \langle x_{i} | e^{-H T /\hbar} | x_{j} \rangle ~ .
\ee
The matrices $U$ and $D$ have the following meaning,
\bea
\label{eq:UDInterpret}
&& U^{\dagger}_{ik}=<x_i|E_k^{eff}> ~ ,
\nonumber \\
&& D_k(T)=e^{-{E_k^{eff}}T/\hbar} ~ .
\eea
The eigenvector $|E_k^{eff}>$ of the effective Hamiltonian $H_{eff}$ can be identified with the column $k$ of matrix $U^{\dagger}$ and
the energy eigenvalues $E^{eff}_{k}$ of $H_{eff}$ can be identified with the logarithm of the diagonal matrix elements of $D(T)$.
Thus via the factorisation (\ref{eq:Factor}) one can define an effective Hamiltonian, 
\be
H_{eff} = \sum_{k =1}^{N} | E^{eff}_{k} > E^{eff}_{k} < E^{eff}_{k} |.
\ee
Note that in the above we have been mathematically a bit sloppy.
The states $| x_{i} \rangle$ are not Hilbert states. We have to replace  
$| x_{i} \rangle$ by some ``localized" Hilbert state. This can be done by introducing box states. We associate to each $x_{i}$ some box state $b_{i}$,
defined by
\be
b_{i}(x) =  \left\{ 
\begin{array}{l}
1/\sqrt{\Delta x_{i}} ~~ \mbox{if} ~~ x_{i} < x \leq x_{i+1}
\\
0 ~~ \mbox{else}
\end{array}
\right.
\ee
where $\Delta x_{i} = x_{i+1}-x_{i}$.
When we use an equidistant distribution of $x_{i}$, i.e. 
$\Delta x_{i} = \Delta x$, we refer to the basis of box states as the regular basis. Below we will consider a stochastic basis.

\section{Matrix elements}
\label{sec:MatElem}
We compute the matrix element $M_{ij}(T)$ directly from 
the action via Monte Carlo with importance sampling. 
This is done by writing each transition matrix element 
as a ratio of two path integrals.
This is done by splitting the action into a non-interacting part and an interacting part,
\be
S = S_{0} + S_{V} ~ .
\ee
This allows to express the transition amplitude (\ref{eq:TransAmpl}) by
\be
M_{ij}(T) = M^{(0)}_{ij}(T) ~
\frac{ 
\left.
\int [dx] ~ \exp[ - S_{V}[x]/\hbar ] ~ \exp[ -S_{0}[x]/\hbar ] \right|^{x_i,T}_{x_j,0} }
{ \left.
\int [dx] ~ \exp[ -S_{0}[x]/\hbar ] \right|^{x_i,T}_{x_j,0} } ~ .
\ee
Here $e^{-S_{0}/\hbar}$ is the weight factor and $e^{-S_{V}/\hbar}$ is the observable.
$M^{(0)}_{ij}(T)$ stands for the transition amplitude of the noninteracting system, which is known analytically.
For details see ref.\cite{Jirari:99}.
Carrying out these steps allows us to construct an effective Hamiltonian, which reproduces well low energy physics, provided that the nodes $x_i$ cover a large enough interval (depending on the range of the potential) and the resolution $\Delta x$ is small enough. This can be achieved with a small numerical effort for a one-body system in 1 dimension.
Our goal is, however, to solve many-body systems. 
What to do in such case is the subject of the following section.

\section{Stochastic basis}
\label{sec:Stochastic}
It is evident that the regular basis defined above becomes prohibitively large
when applied to a many-body system. For example, in a spin model of a 1-dimensional chain of 30 atoms with spin 1/2, the Hilbert space has the dimension $D=2^{30} = 1 073 741 824$.
For such situations we wish to construct a smaller basis which gives an effective Hamiltonian reproducing well low-energy observables.
Why should such a basis exist in the first place?
The heuristic argument is that the Euclidean path integral, when evaluated 
via Monte Carlo with importance sampling, gives a good answer for the 
transition amplitude. In particular, this is possible 
by taking into account a ``small" number of configurations (e.g. on the order of $N_{stoch}=100 - 1000$). 
In a crude way the configurations correspond to basis functions. Thus we expect that suitably chosen basis functions exist, the number of which is in the order of 100 - 1000, which yields a satisfactory effective low energy Hamiltonian.

We shall construct a small basis in a stochastic way, using Monte Carlo.  
We proceed by the following steps:
(i) Compute the Euclidean Green's function $G_{E}(x,T;0,0)$. We define $P(x)$ as the probability density (normalized to unity) obtained  from $G_{E}(x,T;0,0)$. (ii) Find an algorithm giving a random variable $x$ distributed according to $P(x)$ and draw samples from this distribution, giving nodes, say $x_{\nu}$. Finally, one obtains the stochastic basis by constructing the 
corresponding box states from the nodes $x_{\nu}$.  
This goal can be achieved in an elegant and efficient manner via 
the Euclidean path integral, expressing $P(x)$ as 
\bea
\label{eq:ProbDensPathInt}
P(x) = \frac{ \int [dy] \exp[ - S_{E}[y]/\hbar ]\bigg|_{0,0}^{x,T} }
{ \int_{-\infty}^{+\infty} dx 
\int [dy] \exp[ - S_{E}[y]/\hbar ]\bigg|_{0,0}^{x,T} } ~~~ .
\eea
Using a Monte Carlo algorithm with importance sampling (e.g., Metropolis \cite{Metropolis:53}) one generates representative paths, which all start at $x=0$, $t=0$ and arrive 
at some position $x$ at time $t=T$.
Let us denote those paths (configurations) by $C_{\nu} \equiv x_{\nu}(t)$.
We denote the endpoint of path $C_{\nu}$ at time $t=T$ by 
$x_{\nu} \equiv x_{\nu}(T)$.
Those form the stochastically selected nodes, which define the stochastic basis.
One should note that the Green's function $G_{E}$, the probability density $P$
and hence the stochastic basis all depend on the choice of the parameter $T$, related to the inverse temperature $\beta$ via $\beta = T/\hbar$.
It turns out that $T$ influences the size of the finite temperature window in thermodynamical observables. However, it is not the only parameter to do so, 
as this window also depends on the size of the basis.

\section{Numerical results}
\label{sec:Results}
\subsection{Quantum mechanics}
\label{sec:QM} 
The Monte Carlo Hamiltonian has been tested on a number of quantum mechanical potential models, by computing the spectrum, wave functions and thermodynamical observables 
in some low temperature window \cite{Jirari:99,Luo:00,Huang:00,Jiang:00}.
Although the results from the Monte Carlo Hamiltonian agree well with the 
exact results, low-dimensional models in Q.M. are not the target of this method. The target are rather high-dimensional models or models with a large number of degrees of freedom. Our expectation that the MC Hamiltonian works well in such system is guided by the analogy of numerical integration with Monte Carlo:
The Monte Carlo method is {\it not} the most efficient method when doing low-dimensional integrals. However, it is most efficient when doing integrals in high dimensions (e.g. path integrals).

\subsection{Klein-Gordon model}
\label{sec:KleinGord}
We consider in D=1 a chain of coupled harmonic oscillators, which is equivalent to the Klein-Gordon field on a $1+1$ lattice.
The model is given by
\bea
\label{eq:Chain}
S &=& \int dt ~ T - V
\nonumber \\
T &=& \sum_{n=1}^{N} \frac{1}{2} m \dot{q}_{n}^{2}
\nonumber \\
V &=& \frac{1}{2} \sum_{n=1}^{N} \Omega^{2}(q_{n} - q_{n+1})^{2} + \Omega_{0}^{2} q_{n}^{2} ~ .
\eea
The parameters have been chosen as
$m=1$, $\Omega=1$, $\Omega_0=2$,
$N_{\rm{osc}}=9$, $a=1$, $T=2$ and $\hbar=1$.
After the stochastic basis with $N_{stoch}$ is generated,
we obtain the matrix elements $M_{n'n}$
Then we compute the eigenvalues and eigenvectors
using the method described in Sect. (\ref{sec:MCHamiltonian}).
Table [\ref{tab.1}] gives a comparison between the spectrum
from the effective Hamiltonian with the stochastic basis
and the analytic result for the first 20 states. 
We display the results for $N_{stoch} = 10, 100, 1000$ and observe good 
agreement for large enough $N_{stoch}$.
This means that the stochastic basis works well.
\begin{table}[hbt]
\caption{Spectrum of Klein-Gordon model on the lattice,
MC Hamiltonian (stochastic basis) versus
exact result.}
\vspace{3mm}
\begin{center}
\begin{tabular}{|c|c|c|c|c|}
\hline
$n$ & $E_{n}^{\rm{eff}}$ ($N_{stoch}=10$) & $E_{n}^{\rm{eff}}$ ($N_{stoch}=100$) & $E_{n}^{\rm{eff}}$ ($N_{stoch}=1000$)  & $E_{n}^{\rm{exact}}$\\
\hline
   1   & 10.671 & 10.960 &  10.905  &  10.944 \\
   2   & 12.725 & 12.964 &  12.957  &  12.944 \\
   3   & 13.105 & 13.109 &  12.985  &  13.057 \\
   4   & 13.186 & 13.198 &  13.044  &  13.057 \\
   5   & 13.380 & 13.395 &  13.300  &  13.322 \\
   6   & 13.710 & 13.466 &  13.345  &  13.322 \\
   7   & 13.810 & 13.483 &  13.552  &  13.590 \\
   8   & 13.959 & 13.717 &  13.586  &  13.590 \\
   9   & 14.448 & 13.825 &  13.680  &  13.751 \\
   10  & 14.719 & 14.042 &  13.745  &  13.751 \\
   11  &  -     & 14.905 &  14.985  &  14.944 \\
   12  &  -     & 15.105 &  15.012  &  15.058 \\
   13  &  -     & 15.129 &  15.057  &  15.058 \\
   14  &  -     & 15.147 &  15.109  &  15.172 \\
   15  &  -     & 15.287 &  15.125  &  15.172 \\
   16  &  -     & 15.381 &  15.187  &  15.172 \\
   17  &  -     & 15.439 &  15.309  &  15.322 \\
   18  &  -     & 15.451 &  15.396  &  15.322 \\
   19  &  -     & 15.496 &  15.421  &  15.435 \\
   20  &  -     & 15.496 &  15.433  &  15.435 \\
\hline
\end{tabular}
\end{center}
\vspace{0mm}
\label{tab.1}
\end{table}
\\

\noindent We have also computed thermodynamical quantities
such as the partition function $Z$, free energy $F$,
average energy $U=\overline{E}$ and specific heat $C$.
The analytical results are
\begin{eqnarray}
Z(\beta) &=& {\rm Tr} \left( \exp \left( -\beta H \right) \right)
=\prod_{l=1}^{N_{\rm{osc}}} {1 \over 2 {\rm sinh} \left( \beta \hbar
\omega_l/2 \right)},
\nonumber \\
\overline{E}(\beta) &=& {1 \over Z} {\rm Tr} \left(H \exp \left( -\beta H
\right) \right)
= - {\partial \log Z \over \partial \beta}
= \sum_{l=1}^{N_{\rm{osc}}} {\hbar \omega_l \over 2} \coth \left( \beta
\hbar \omega_l/2 \right),
\nonumber \\
C(\beta) &=& \frac{\partial \overline{E}}{\partial {\it \tau}}
= -k_B \beta^2 \frac{\partial \overline{E}}{\partial \beta}
= k_B \sum_{l=1}^{N_{\rm{osc}}}
\left( {\beta \hbar \omega_l/2 \over 2 {\rm sinh}
\left( \beta \hbar \omega_l/2 \right)} \right)^2 ,
\end{eqnarray}
where 
\begin{eqnarray}
\omega_l &=& \sqrt{\Omega_0^2+4\Omega^2 sin^2(p_l \Delta x/2)} ~ ,
\nonumber \\
\Delta p &=& 2 \pi / (N_{\rm{osc}} \Delta x) ~ ,
\nonumber \\
x_j &=& [-(N_{\rm{osc}} -1) / 2 + (j-1)]\Delta x ~ ,
\nonumber \\
p_l &=& [-(N_{\rm{osc}} -1) / 2 + (l-1)]\Delta p ~ .
\end{eqnarray}
Here $j$ and $l$ run from 1 to $N_{\rm{osc}}$ (number of oscillators).
$\Delta x=a=1$ is the lattice spacing, $\beta=T/\hbar$, 
the temperature is related to $\beta$ via ${\it \tau} = 1/(\beta k_B)$, 
and $k_B$ is the Boltzmann constant.

\begin{figure}[htb]
%\vspace{-4mm}
%\vspace{-12mm}
\begin{center}
% GNUPLOT: LaTeX picture
\setlength{\unitlength}{0.240900pt}
\ifx\plotpoint\undefined\newsavebox{\plotpoint}\fi
\sbox{\plotpoint}{\rule[-0.200pt]{0.400pt}{0.400pt}}%
\begin{picture}(1349,809)(0,0)
\font\gnuplot=cmr10 at 10pt
\gnuplot
\sbox{\plotpoint}{\rule[-0.200pt]{0.400pt}{0.400pt}}%
\put(141.0,123.0){\rule[-0.200pt]{4.818pt}{0.400pt}}
\put(121,123){\makebox(0,0)[r]{10}}
\put(1308.0,123.0){\rule[-0.200pt]{4.818pt}{0.400pt}}
\put(141.0,182.0){\rule[-0.200pt]{4.818pt}{0.400pt}}
\put(121,182){\makebox(0,0)[r]{11}}
\put(1308.0,182.0){\rule[-0.200pt]{4.818pt}{0.400pt}}
\put(141.0,240.0){\rule[-0.200pt]{4.818pt}{0.400pt}}
\put(121,240){\makebox(0,0)[r]{12}}
\put(1308.0,240.0){\rule[-0.200pt]{4.818pt}{0.400pt}}
\put(141.0,299.0){\rule[-0.200pt]{4.818pt}{0.400pt}}
\put(121,299){\makebox(0,0)[r]{13}}
\put(1308.0,299.0){\rule[-0.200pt]{4.818pt}{0.400pt}}
\put(141.0,358.0){\rule[-0.200pt]{4.818pt}{0.400pt}}
\put(121,358){\makebox(0,0)[r]{14}}
\put(1308.0,358.0){\rule[-0.200pt]{4.818pt}{0.400pt}}
\put(141.0,417.0){\rule[-0.200pt]{4.818pt}{0.400pt}}
\put(121,417){\makebox(0,0)[r]{15}}
\put(1308.0,417.0){\rule[-0.200pt]{4.818pt}{0.400pt}}
\put(141.0,475.0){\rule[-0.200pt]{4.818pt}{0.400pt}}
\put(121,475){\makebox(0,0)[r]{16}}
\put(1308.0,475.0){\rule[-0.200pt]{4.818pt}{0.400pt}}
\put(141.0,534.0){\rule[-0.200pt]{4.818pt}{0.400pt}}
\put(121,534){\makebox(0,0)[r]{17}}
\put(1308.0,534.0){\rule[-0.200pt]{4.818pt}{0.400pt}}
\put(141.0,593.0){\rule[-0.200pt]{4.818pt}{0.400pt}}
\put(121,593){\makebox(0,0)[r]{18}}
\put(1308.0,593.0){\rule[-0.200pt]{4.818pt}{0.400pt}}
\put(141.0,652.0){\rule[-0.200pt]{4.818pt}{0.400pt}}
\put(121,652){\makebox(0,0)[r]{19}}
\put(1308.0,652.0){\rule[-0.200pt]{4.818pt}{0.400pt}}
\put(141.0,710.0){\rule[-0.200pt]{4.818pt}{0.400pt}}
\put(121,710){\makebox(0,0)[r]{20}}
\put(1308.0,710.0){\rule[-0.200pt]{4.818pt}{0.400pt}}
\put(141.0,769.0){\rule[-0.200pt]{4.818pt}{0.400pt}}
\put(121,769){\makebox(0,0)[r]{21}}
\put(1308.0,769.0){\rule[-0.200pt]{4.818pt}{0.400pt}}
\put(141.0,123.0){\rule[-0.200pt]{0.400pt}{4.818pt}}
\put(141,82){\makebox(0,0){0}}
\put(141.0,749.0){\rule[-0.200pt]{0.400pt}{4.818pt}}
\put(260.0,123.0){\rule[-0.200pt]{0.400pt}{4.818pt}}
\put(260,82){\makebox(0,0){1}}
\put(260.0,749.0){\rule[-0.200pt]{0.400pt}{4.818pt}}
\put(378.0,123.0){\rule[-0.200pt]{0.400pt}{4.818pt}}
\put(378,82){\makebox(0,0){2}}
\put(378.0,749.0){\rule[-0.200pt]{0.400pt}{4.818pt}}
\put(497.0,123.0){\rule[-0.200pt]{0.400pt}{4.818pt}}
\put(497,82){\makebox(0,0){3}}
\put(497.0,749.0){\rule[-0.200pt]{0.400pt}{4.818pt}}
\put(616.0,123.0){\rule[-0.200pt]{0.400pt}{4.818pt}}
\put(616,82){\makebox(0,0){4}}
\put(616.0,749.0){\rule[-0.200pt]{0.400pt}{4.818pt}}
\put(735.0,123.0){\rule[-0.200pt]{0.400pt}{4.818pt}}
\put(735,82){\makebox(0,0){5}}
\put(735.0,749.0){\rule[-0.200pt]{0.400pt}{4.818pt}}
\put(853.0,123.0){\rule[-0.200pt]{0.400pt}{4.818pt}}
\put(853,82){\makebox(0,0){6}}
\put(853.0,749.0){\rule[-0.200pt]{0.400pt}{4.818pt}}
\put(972.0,123.0){\rule[-0.200pt]{0.400pt}{4.818pt}}
\put(972,82){\makebox(0,0){7}}
\put(972.0,749.0){\rule[-0.200pt]{0.400pt}{4.818pt}}
\put(1091.0,123.0){\rule[-0.200pt]{0.400pt}{4.818pt}}
\put(1091,82){\makebox(0,0){8}}
\put(1091.0,749.0){\rule[-0.200pt]{0.400pt}{4.818pt}}
\put(1209.0,123.0){\rule[-0.200pt]{0.400pt}{4.818pt}}
\put(1209,82){\makebox(0,0){9}}
\put(1209.0,749.0){\rule[-0.200pt]{0.400pt}{4.818pt}}
\put(1328.0,123.0){\rule[-0.200pt]{0.400pt}{4.818pt}}
\put(1328,82){\makebox(0,0){10}}
\put(1328.0,749.0){\rule[-0.200pt]{0.400pt}{4.818pt}}
\put(141.0,123.0){\rule[-0.200pt]{285.948pt}{0.400pt}}
\put(1328.0,123.0){\rule[-0.200pt]{0.400pt}{155.621pt}}
\put(141.0,769.0){\rule[-0.200pt]{285.948pt}{0.400pt}}
\put(30,446){\makebox(0,0){$\overline{E}$}}
\put(734,21){\makebox(0,0){$\beta$}}
\put(141.0,123.0){\rule[-0.200pt]{0.400pt}{155.621pt}}
\put(1156,729){\makebox(0,0)[r]{MC Hamiltonian}}
\put(200,534){\raisebox{-.8pt}{\makebox(0,0){$\Delta$}}}
\put(260,296){\raisebox{-.8pt}{\makebox(0,0){$\Delta$}}}
\put(319,211){\raisebox{-.8pt}{\makebox(0,0){$\Delta$}}}
\put(378,187){\raisebox{-.8pt}{\makebox(0,0){$\Delta$}}}
\put(438,179){\raisebox{-.8pt}{\makebox(0,0){$\Delta$}}}
\put(497,177){\raisebox{-.8pt}{\makebox(0,0){$\Delta$}}}
\put(556,176){\raisebox{-.8pt}{\makebox(0,0){$\Delta$}}}
\put(616,176){\raisebox{-.8pt}{\makebox(0,0){$\Delta$}}}
\put(675,176){\raisebox{-.8pt}{\makebox(0,0){$\Delta$}}}
\put(735,176){\raisebox{-.8pt}{\makebox(0,0){$\Delta$}}}
\put(794,176){\raisebox{-.8pt}{\makebox(0,0){$\Delta$}}}
\put(853,176){\raisebox{-.8pt}{\makebox(0,0){$\Delta$}}}
\put(913,176){\raisebox{-.8pt}{\makebox(0,0){$\Delta$}}}
\put(972,176){\raisebox{-.8pt}{\makebox(0,0){$\Delta$}}}
\put(1031,176){\raisebox{-.8pt}{\makebox(0,0){$\Delta$}}}
\put(1091,176){\raisebox{-.8pt}{\makebox(0,0){$\Delta$}}}
\put(1150,176){\raisebox{-.8pt}{\makebox(0,0){$\Delta$}}}
\put(1209,176){\raisebox{-.8pt}{\makebox(0,0){$\Delta$}}}
\put(1269,176){\raisebox{-.8pt}{\makebox(0,0){$\Delta$}}}
\put(1328,176){\raisebox{-.8pt}{\makebox(0,0){$\Delta$}}}
\put(1232,729){\raisebox{-.8pt}{\makebox(0,0){$\Delta$}}}
\put(1156,688){\makebox(0,0)[r]{Analytical result}}
\multiput(1176,688)(20.756,0.000){6}{\usebox{\plotpoint}}
\put(1288,688){\usebox{\plotpoint}}
\put(200,722){\usebox{\plotpoint}}
\multiput(200,722)(2.574,-20.595){5}{\usebox{\plotpoint}}
\multiput(212,626)(2.464,-20.609){5}{\usebox{\plotpoint}}
\multiput(223,534)(2.901,-20.552){4}{\usebox{\plotpoint}}
\multiput(235,449)(3.052,-20.530){3}{\usebox{\plotpoint}}
\multiput(246,375)(3.743,-20.415){3}{\usebox{\plotpoint}}
\multiput(257,315)(5.579,-19.992){2}{\usebox{\plotpoint}}
\multiput(269,272)(7.589,-19.318){2}{\usebox{\plotpoint}}
\put(287.00,233.83){\usebox{\plotpoint}}
\put(302.39,220.41){\usebox{\plotpoint}}
\put(321.89,213.37){\usebox{\plotpoint}}
\put(341.14,205.74){\usebox{\plotpoint}}
\put(359.95,197.02){\usebox{\plotpoint}}
\put(379.17,189.28){\usebox{\plotpoint}}
\put(399.45,185.01){\usebox{\plotpoint}}
\put(420.06,182.72){\usebox{\plotpoint}}
\put(440.78,181.93){\usebox{\plotpoint}}
\put(461.49,181.00){\usebox{\plotpoint}}
\put(482.21,180.00){\usebox{\plotpoint}}
\put(502.94,179.46){\usebox{\plotpoint}}
\put(523.67,179.00){\usebox{\plotpoint}}
\put(544.43,179.00){\usebox{\plotpoint}}
\put(565.19,179.00){\usebox{\plotpoint}}
\put(585.94,179.00){\usebox{\plotpoint}}
\put(606.70,179.00){\usebox{\plotpoint}}
\put(627.45,179.00){\usebox{\plotpoint}}
\put(648.19,178.71){\usebox{\plotpoint}}
\put(668.92,178.00){\usebox{\plotpoint}}
\put(689.67,178.00){\usebox{\plotpoint}}
\put(710.43,178.00){\usebox{\plotpoint}}
\put(731.18,178.00){\usebox{\plotpoint}}
\put(751.94,178.00){\usebox{\plotpoint}}
\put(772.69,178.00){\usebox{\plotpoint}}
\put(793.45,178.00){\usebox{\plotpoint}}
\put(814.21,178.00){\usebox{\plotpoint}}
\put(834.96,178.00){\usebox{\plotpoint}}
\put(855.72,178.00){\usebox{\plotpoint}}
\put(876.47,178.00){\usebox{\plotpoint}}
\put(897.23,178.00){\usebox{\plotpoint}}
\put(917.98,178.00){\usebox{\plotpoint}}
\put(938.74,178.00){\usebox{\plotpoint}}
\put(959.49,178.00){\usebox{\plotpoint}}
\put(980.25,178.00){\usebox{\plotpoint}}
\put(1001.01,178.00){\usebox{\plotpoint}}
\put(1021.76,178.00){\usebox{\plotpoint}}
\put(1042.52,178.00){\usebox{\plotpoint}}
\put(1063.27,178.00){\usebox{\plotpoint}}
\put(1084.03,178.00){\usebox{\plotpoint}}
\put(1104.78,178.00){\usebox{\plotpoint}}
\put(1125.54,178.00){\usebox{\plotpoint}}
\put(1146.29,178.00){\usebox{\plotpoint}}
\put(1167.05,178.00){\usebox{\plotpoint}}
\put(1187.80,178.00){\usebox{\plotpoint}}
\put(1208.56,178.00){\usebox{\plotpoint}}
\put(1229.32,178.00){\usebox{\plotpoint}}
\put(1250.07,178.00){\usebox{\plotpoint}}
\put(1270.83,178.00){\usebox{\plotpoint}}
\put(1291.58,178.00){\usebox{\plotpoint}}
\put(1312.34,178.00){\usebox{\plotpoint}}
\put(1328,178){\usebox{\plotpoint}}
\end{picture}
\end{center}
\caption{Average energy $U$ of the Klein-Gordon model on a 1+1 dimensional
lattice.}
\label{fig.1}
%\vspace{-2mm}
\end{figure}
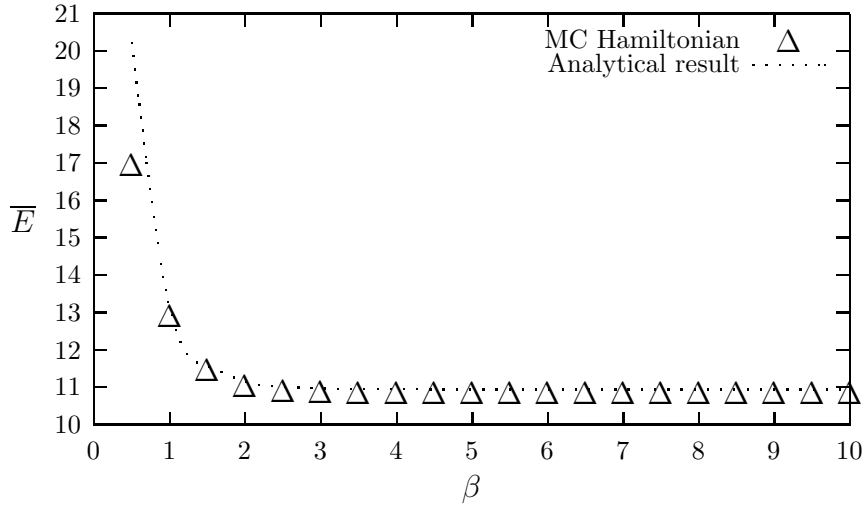

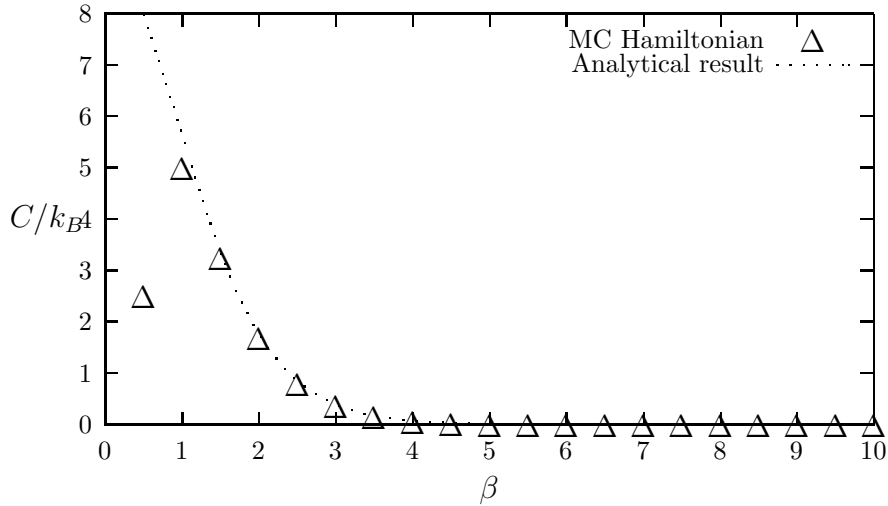
\begin{figure}[htb]
%\vspace{-4mm}
%\vspace{-12mm}
\begin{center}
% GNUPLOT: LaTeX picture
\setlength{\unitlength}{0.240900pt}
\ifx\plotpoint\undefined\newsavebox{\plotpoint}\fi
\sbox{\plotpoint}{\rule[-0.200pt]{0.400pt}{0.400pt}}%
\begin{picture}(1349,809)(0,0)
\font\gnuplot=cmr10 at 10pt
\gnuplot
\sbox{\plotpoint}{\rule[-0.200pt]{0.400pt}{0.400pt}}%
\put(121.0,123.0){\rule[-0.200pt]{4.818pt}{0.400pt}}
\put(101,123){\makebox(0,0)[r]{0}}
\put(1308.0,123.0){\rule[-0.200pt]{4.818pt}{0.400pt}}
\put(121.0,204.0){\rule[-0.200pt]{4.818pt}{0.400pt}}
\put(101,204){\makebox(0,0)[r]{1}}
\put(1308.0,204.0){\rule[-0.200pt]{4.818pt}{0.400pt}}
\put(121.0,285.0){\rule[-0.200pt]{4.818pt}{0.400pt}}
\put(101,285){\makebox(0,0)[r]{2}}
\put(1308.0,285.0){\rule[-0.200pt]{4.818pt}{0.400pt}}
\put(121.0,365.0){\rule[-0.200pt]{4.818pt}{0.400pt}}
\put(101,365){\makebox(0,0)[r]{3}}
\put(1308.0,365.0){\rule[-0.200pt]{4.818pt}{0.400pt}}
\put(121.0,446.0){\rule[-0.200pt]{4.818pt}{0.400pt}}
\put(101,446){\makebox(0,0)[r]{4}}
\put(1308.0,446.0){\rule[-0.200pt]{4.818pt}{0.400pt}}
\put(121.0,527.0){\rule[-0.200pt]{4.818pt}{0.400pt}}
\put(101,527){\makebox(0,0)[r]{5}}
\put(1308.0,527.0){\rule[-0.200pt]{4.818pt}{0.400pt}}
\put(121.0,608.0){\rule[-0.200pt]{4.818pt}{0.400pt}}
\put(101,608){\makebox(0,0)[r]{6}}
\put(1308.0,608.0){\rule[-0.200pt]{4.818pt}{0.400pt}}
\put(121.0,688.0){\rule[-0.200pt]{4.818pt}{0.400pt}}
\put(101,688){\makebox(0,0)[r]{7}}
\put(1308.0,688.0){\rule[-0.200pt]{4.818pt}{0.400pt}}
\put(121.0,769.0){\rule[-0.200pt]{4.818pt}{0.400pt}}
\put(101,769){\makebox(0,0)[r]{8}}
\put(1308.0,769.0){\rule[-0.200pt]{4.818pt}{0.400pt}}
\put(121.0,123.0){\rule[-0.200pt]{0.400pt}{4.818pt}}
\put(121,82){\makebox(0,0){0}}
\put(121.0,749.0){\rule[-0.200pt]{0.400pt}{4.818pt}}
\put(242.0,123.0){\rule[-0.200pt]{0.400pt}{4.818pt}}
\put(242,82){\makebox(0,0){1}}
\put(242.0,749.0){\rule[-0.200pt]{0.400pt}{4.818pt}}
\put(362.0,123.0){\rule[-0.200pt]{0.400pt}{4.818pt}}
\put(362,82){\makebox(0,0){2}}
\put(362.0,749.0){\rule[-0.200pt]{0.400pt}{4.818pt}}
\put(483.0,123.0){\rule[-0.200pt]{0.400pt}{4.818pt}}
\put(483,82){\makebox(0,0){3}}
\put(483.0,749.0){\rule[-0.200pt]{0.400pt}{4.818pt}}
\put(604.0,123.0){\rule[-0.200pt]{0.400pt}{4.818pt}}
\put(604,82){\makebox(0,0){4}}
\put(604.0,749.0){\rule[-0.200pt]{0.400pt}{4.818pt}}
\put(725.0,123.0){\rule[-0.200pt]{0.400pt}{4.818pt}}
\put(725,82){\makebox(0,0){5}}
\put(725.0,749.0){\rule[-0.200pt]{0.400pt}{4.818pt}}
\put(845.0,123.0){\rule[-0.200pt]{0.400pt}{4.818pt}}
\put(845,82){\makebox(0,0){6}}
\put(845.0,749.0){\rule[-0.200pt]{0.400pt}{4.818pt}}
\put(966.0,123.0){\rule[-0.200pt]{0.400pt}{4.818pt}}
\put(966,82){\makebox(0,0){7}}
\put(966.0,749.0){\rule[-0.200pt]{0.400pt}{4.818pt}}
\put(1087.0,123.0){\rule[-0.200pt]{0.400pt}{4.818pt}}
\put(1087,82){\makebox(0,0){8}}
\put(1087.0,749.0){\rule[-0.200pt]{0.400pt}{4.818pt}}
\put(1207.0,123.0){\rule[-0.200pt]{0.400pt}{4.818pt}}
\put(1207,82){\makebox(0,0){9}}
\put(1207.0,749.0){\rule[-0.200pt]{0.400pt}{4.818pt}}
\put(1328.0,123.0){\rule[-0.200pt]{0.400pt}{4.818pt}}
\put(1328,82){\makebox(0,0){10}}
\put(1328.0,749.0){\rule[-0.200pt]{0.400pt}{4.818pt}}
\put(121.0,123.0){\rule[-0.200pt]{290.766pt}{0.400pt}}
\put(1328.0,123.0){\rule[-0.200pt]{0.400pt}{155.621pt}}
\put(121.0,769.0){\rule[-0.200pt]{290.766pt}{0.400pt}}
\put(30,446){\makebox(0,0){$C/k_B$}}
\put(724,21){\makebox(0,0){$\beta$}}
\put(121.0,123.0){\rule[-0.200pt]{0.400pt}{155.621pt}}
\put(1156,729){\makebox(0,0)[r]{MC Hamiltonian}}
\put(181,327){\raisebox{-.8pt}{\makebox(0,0){$\Delta$}}}
\put(242,528){\raisebox{-.8pt}{\makebox(0,0){$\Delta$}}}
\put(302,386){\raisebox{-.8pt}{\makebox(0,0){$\Delta$}}}
\put(362,261){\raisebox{-.8pt}{\makebox(0,0){$\Delta$}}}
\put(423,189){\raisebox{-.8pt}{\makebox(0,0){$\Delta$}}}
\put(483,153){\raisebox{-.8pt}{\makebox(0,0){$\Delta$}}}
\put(543,136){\raisebox{-.8pt}{\makebox(0,0){$\Delta$}}}
\put(604,129){\raisebox{-.8pt}{\makebox(0,0){$\Delta$}}}
\put(664,125){\raisebox{-.8pt}{\makebox(0,0){$\Delta$}}}
\put(725,124){\raisebox{-.8pt}{\makebox(0,0){$\Delta$}}}
\put(785,123){\raisebox{-.8pt}{\makebox(0,0){$\Delta$}}}
\put(845,123){\raisebox{-.8pt}{\makebox(0,0){$\Delta$}}}
\put(906,123){\raisebox{-.8pt}{\makebox(0,0){$\Delta$}}}
\put(966,123){\raisebox{-.8pt}{\makebox(0,0){$\Delta$}}}
\put(1026,123){\raisebox{-.8pt}{\makebox(0,0){$\Delta$}}}
\put(1087,123){\raisebox{-.8pt}{\makebox(0,0){$\Delta$}}}
\put(1147,123){\raisebox{-.8pt}{\makebox(0,0){$\Delta$}}}
\put(1207,123){\raisebox{-.8pt}{\makebox(0,0){$\Delta$}}}
\put(1268,123){\raisebox{-.8pt}{\makebox(0,0){$\Delta$}}}
\put(1328,123){\raisebox{-.8pt}{\makebox(0,0){$\Delta$}}}
\put(1232,729){\raisebox{-.8pt}{\makebox(0,0){$\Delta$}}}
\put(1156,688){\makebox(0,0)[r]{Analytical result}}
\multiput(1176,688)(20.756,0.000){6}{\usebox{\plotpoint}}
\put(1288,688){\usebox{\plotpoint}}
\put(181,766){\usebox{\plotpoint}}
\multiput(181,766)(6.563,-19.690){2}{\usebox{\plotpoint}}
\multiput(193,730)(6.563,-19.690){2}{\usebox{\plotpoint}}
\multiput(205,694)(6.065,-19.850){2}{\usebox{\plotpoint}}
\multiput(216,658)(6.563,-19.690){2}{\usebox{\plotpoint}}
\multiput(228,622)(5.915,-19.895){2}{\usebox{\plotpoint}}
\multiput(239,585)(6.403,-19.743){2}{\usebox{\plotpoint}}
\put(256.72,528.75){\usebox{\plotpoint}}
\multiput(262,511)(6.403,-19.743){2}{\usebox{\plotpoint}}
\multiput(274,474)(6.732,-19.634){2}{\usebox{\plotpoint}}
\multiput(286,439)(6.563,-19.690){2}{\usebox{\plotpoint}}
\put(302.81,391.00){\usebox{\plotpoint}}
\multiput(309,375)(7.589,-19.318){2}{\usebox{\plotpoint}}
\put(326.30,333.35){\usebox{\plotpoint}}
\put(335.31,314.66){\usebox{\plotpoint}}
\multiput(344,298)(9.631,-18.386){2}{\usebox{\plotpoint}}
\put(365.56,260.29){\usebox{\plotpoint}}
\put(376.81,242.84){\usebox{\plotpoint}}
\put(389.59,226.51){\usebox{\plotpoint}}
\put(402.63,210.37){\usebox{\plotpoint}}
\put(417.67,196.10){\usebox{\plotpoint}}
\put(433.68,182.90){\usebox{\plotpoint}}
\put(450.82,171.21){\usebox{\plotpoint}}
\put(468.89,161.05){\usebox{\plotpoint}}
\put(487.83,152.57){\usebox{\plotpoint}}
\put(507.37,145.63){\usebox{\plotpoint}}
\put(527.45,140.39){\usebox{\plotpoint}}
\put(547.74,136.06){\usebox{\plotpoint}}
\put(568.25,133.23){\usebox{\plotpoint}}
\put(588.83,130.67){\usebox{\plotpoint}}
\put(609.43,129.00){\usebox{\plotpoint}}
\put(630.11,127.24){\usebox{\plotpoint}}
\put(650.82,126.00){\usebox{\plotpoint}}
\put(671.53,125.00){\usebox{\plotpoint}}
\put(692.28,124.89){\usebox{\plotpoint}}
\put(713.00,124.00){\usebox{\plotpoint}}
\put(733.75,124.00){\usebox{\plotpoint}}
\put(754.51,124.00){\usebox{\plotpoint}}
\put(775.25,123.73){\usebox{\plotpoint}}
\put(795.98,123.00){\usebox{\plotpoint}}
\put(816.73,123.00){\usebox{\plotpoint}}
\put(837.49,123.00){\usebox{\plotpoint}}
\put(858.25,123.00){\usebox{\plotpoint}}
\put(879.00,123.00){\usebox{\plotpoint}}
\put(899.76,123.00){\usebox{\plotpoint}}
\put(920.51,123.00){\usebox{\plotpoint}}
\put(941.27,123.00){\usebox{\plotpoint}}
\put(962.02,123.00){\usebox{\plotpoint}}
\put(982.78,123.00){\usebox{\plotpoint}}
\put(1003.53,123.00){\usebox{\plotpoint}}
\put(1024.29,123.00){\usebox{\plotpoint}}
\put(1045.05,123.00){\usebox{\plotpoint}}
\put(1065.80,123.00){\usebox{\plotpoint}}
\put(1086.56,123.00){\usebox{\plotpoint}}
\put(1107.31,123.00){\usebox{\plotpoint}}
\put(1128.07,123.00){\usebox{\plotpoint}}
\put(1148.82,123.00){\usebox{\plotpoint}}
\put(1169.58,123.00){\usebox{\plotpoint}}
\put(1190.33,123.00){\usebox{\plotpoint}}
\put(1211.09,123.00){\usebox{\plotpoint}}
\put(1231.84,123.00){\usebox{\plotpoint}}
\put(1252.60,123.00){\usebox{\plotpoint}}
\put(1273.36,123.00){\usebox{\plotpoint}}
\put(1294.11,123.00){\usebox{\plotpoint}}
\put(1314.87,123.00){\usebox{\plotpoint}}
\put(1328,123){\usebox{\plotpoint}}
\end{picture}
\end{center}
\caption{Specific heat ($C/k_B$) of the Klein-Gordon model on a 1+1
dimensional lattice.}
\label{fig.2}
%\vspace{-2mm}
\end{figure}

Since we have approximated $H$ by $H_{\rm{eff}}$,
we can express those thermodynamical observables
via the eigenvalues of the effective Hamiltonian
\begin{eqnarray}
Z^{\rm{eff}}(\beta) &=& \sum_{n=1}^{N}e^{-\beta E_{n}^{\rm{eff}}},
\nonumber \\
\overline{E}^{\rm{eff}}(\beta) &=& \sum_{n=1}^{N}
{E_{n}^{\rm{eff}} {\rm e}^{-\beta E_{n}^{\rm{eff}}} \over
Z^{\rm{eff}}(\beta)},
\nonumber \\
C^{\rm{eff}}(\beta) &=& k_B{\beta}^2\left(\sum_{n=1}^{N}
{(E_{n}^{\rm{eff}})^2e^{-\beta E_{n}^{\rm{eff}}} \over
Z^{\rm{eff}}(\beta)}-\left(\overline{E}^{\rm{eff}}(\beta)\right)^2 \right).
\end{eqnarray}
Here $N$ is the maximum number of eigenvalues available for a given $N_{stoch}$.
Since this is a static system, the eigenvalues should in principle not vary
with $\beta$.
This is observed numerically within statistical errors when $\beta$ and $N_{stoch}$ are not too small.
Using this assumption and
the spectrum at $\beta=2$, we can compute thermodynamical quantities for
other values of $\beta$. For the $N_{stoch}=1000$ case,
the average energy as a function of $\beta$ is shown in Fig.[1] and the specific heat is shown in Fig.[2].
The results from the MC Hamiltonian are in good agreement
with the analytical results for $\beta > 1$. 
However, for $\beta < \beta_{low} \approx 1$, which corresponds to the temperature $\tau > \tau_{up} = (k_{B} \beta_{low})^{-1}$, then the specific heat and also the average energy obtained from the MC Hamiltonian start to deviate from the analytical result. This shows that the MC Hamiltonian is valid only in some finite temperature window.

\section{Summary}

In this paper, we have extended the Monte Carlo Hamiltonian method
with a stochastic basis to quantum field theory (QFT), 
and taken the Klein-Gordon model as an example.
The results are very encouraging. We believe that the application
of the algorithm to more complicated systems will be very interesting. 
It will allow a non-perturbative investigation of physics beyond the ground state.

\vspace{0.5cm}

\noindent {\bf Acknowledgements} \\ 
H.K. and K.M. are grateful for support by NSERC Canada. 
X.Q.L. has been supported by NSF for Distinguished Young Scientists, NSFC, 
Guangdong Provincial NSF, Ministry of Education of China, and Foundation of
Zhongshan Univ. Advanced Research Center.

\end{document}